\documentclass{bmcart}

\RequirePackage{natbib}
\usepackage[utf8]{inputenc} %
\usepackage{booktabs}

\def\includegraphics{}

\startlocaldefs

\endlocaldefs

\usepackage{epsfig,url,placeins}
\usepackage{cleveref}
\usepackage{multirow}
\usepackage{longtable,array,setspace}
\usepackage{rotating} %
\usepackage[font=footnotesize]{caption}
\usepackage{subcaption}
\usepackage[normalem]{ulem} %

\newcommand{\todo}[1]{}
\renewcommand{\todo}[1]{{\color{red}[[TODO: {#1}]]}}

\crefname{section}{Sec.}{Secs.}
\Crefname{section}{Section}{Sections}
\crefname{figure}{Fig.}{Figs.}
\Crefname{figure}{Figure}{Figures}
\crefname{equation}{Eq.}{Eqs.}
\Crefname{equation}{Equation}{Equations}
\crefname{table}{Tab.}{Tabs.}
\Crefname{Table}{Table}{Tables}

\usepackage{comment,soul}
\definecolor{reddish}{HTML}{FBB4AE}
\definecolor{blueish}{HTML}{B3CDE3}
\definecolor{magentish}{HTML}{FF00AA}
\definecolor{greenish}{HTML}{a1d99b}

\usepackage{lineno}
\begin{document}

\begin{frontmatter}

\begin{fmbox}
\dochead{Research}

\title{The penumbra of open source: projects outside of centralized platforms are longer maintained, more academic and more collaborative
}

\author[
   addressref={csds},                   %
   email={milo.trujillo@uvm.edu}   %
]{\inits{MZT}\fnm{Milo Z} \snm{Trujillo}}
\author[
   addressref={csds,cs},
   email={laurent.hebert-dufresne@uvm.edu}
]{\inits{LHD}\fnm{Laurent} \snm{H\'ebert-Dufresne}}
\author[
   addressref={csds,math},
   corref={csds}, %
   email={james.bagrow@uvm.edu}
]{\inits{JPB}\fnm{James} \snm{Bagrow}}

\address[id=csds]{%
  \orgname{Complex Systems Center, University of Vermont}, %
  \city{Burlington, Vermont},                              %
  \cny{USA}                                    %
}
\address[id=cs]{%
  \orgname{Department of Computer Science, University of Vermont}, %
  \city{Burlington, Vermont},                              %
  \cny{USA}                                    %
}
\address[id=math]{%
  \orgname{Department of Mathematics \& Statistics, University of Vermont}, %
  \city{Burlington, Vermont},                              %
  \cny{USA}                                    %
}

\begin{artnotes}
\end{artnotes}

\end{fmbox}%

\begin{abstractbox}

\begin{abstract} %
GitHub has become the central online platform for much of open source, hosting most open source code repositories.
With this popularity, the public digital traces of GitHub are now a valuable means to study teamwork and collaboration.
In many ways, however, GitHub is a convenience sample, and may not be representative of open source development off the platform.
Here we develop a novel, extensive sample of public open source project repositories outside of centralized platforms.
We characterized these projects along a number of dimensions, and compare to a time-matched sample of corresponding GitHub projects.
Our sample projects tend to have more collaborators, are maintained for longer periods, and tend to be more focused on academic and scientific problems.
\end{abstract}


\end{abstractbox}

\end{frontmatter}

\section{Introduction}

The GitHub hosting platform has long been recognized as a promising window into the complex world of online collaborations \cite{github_social_transparency}, open science \cite{perkel2016democratic}, education \cite{zagalsky2015emergence}, public sector work \cite{mergel2015open}, and software development \cite{kalliamvakou2015open}. 
From 10 million repositories in 2014 \cite{promises_perils_github}, GitHub reported over 60 million new repositories in 2020 \cite{2020octoverse}. 
However, despite its size, there remain significant risks associated with GitHub as a data platform \cite{kalliamvakou2016depth}. Without a baseline study examining open source development off of GitHub, it is unclear whether public GitHub data is a representative sample of software development practices or collaborative behavior.
For studies of collaborations, it is particularly worrisome that most GitHub repositories are private and that most public ones are small and inactive \cite{promises_perils_github}. 
These data biases have only grown in recent years as the platform stopped limiting the number of private repositories with fewer than four collaborators in 2019 \cite{newgithub}.

Despite the fact that GitHub is not a transparent or unbiased window into collaborations, the popularity of the platform alone has proved very attractive for researchers.
Early research focused on the value of transparency and working in public, analyzing how individuals select projects and collaborations \cite{github_social_transparency}, and conversely how collaborations grow and thrive \cite{bagrow_teams_2016,muric2019}. 
While fine-grained information about git commits within code repositories is readily available, higher-level findings about team collaboration and social software development practices are scarcer. \citet{bagrow_teams_2016} introduce a metric of ``effective team size,'' measuring each contributor's contributions against the distribution of GitHub events for the repository, distinguishing peripheral and ``drive-by'' contributors from more active team members. \citet{choudhary2018modeling} focus on identifying ``periods of activity'' within a repository, beginning with a simple measurement of time dispersion between GitHub events, then identifying the participants and files edited in each burst of activity to determine productivity and partitioning of work according to apparent team dynamics.

Beyond looking at patterns of collaborations within projects, it is also useful to study GitHub as a social network, where collaborations are social ties mediated by repository \cite{lima2014coding, github_stats, celinska2018coding}.
These studies tend to offer results showing analogies between GitHub collaborations and more classic online social networks, such as modular structure \cite{zoller2020topology} and heterogeneous distributions of collaborators per individual driven by rich-get-richer effects \cite{lima2014coding,celinska2018coding}.
More interestingly, studies also found that GitHub tends to show extremely low levels of reciprocity in actual social connections \cite{lima2014coding} and high levels of hierarchical, often star-like groups \cite{zoller2020topology}.
There are unfortunately few studies providing context for GitHub-specific findings, and no clear baseline to which they should be compared.
Is GitHub more or less collaborative than other platforms of open source development?
How much are collaborations shaped by the open source nature of the work, by the underlying technology, and by the platform itself?
Altogether, it remains an open problem to quantify just how collaborative and social GitHub is. 

GitHub is far from the only platform to host open source projects that use the Git version control system, but it is the most popular.
What remains unclear is how much of the open source ecosystem now exists in GitHub's shadow, and how different these open source projects are when compared to their counterpart on the most popular public platforms.
To this end, here we aim to study what we call the \emph{Penumbra} of open source: Public repositories on public hosts other than the large centralized platforms (e.g.
GitHub, GitLab, Sourceforge and other forges).
Specifically, we want to compare the size, the nature and the temporal patterns of collaborations that occur in the Penumbra with that of a comparable random subset of GitHub.

Open source has long been linked to academic institutions~\cite{lakhani2003hackers}, including libraries~\cite{payne2010,chen2014functionality}, 
research centers~\cite{murphy2012current,pearce2012building},
and the classroom~\cite{van2009adopting}.
Version control systems such as git have been interesting tools to assist in classroom learning~\cite{haaranen2015teaching,clifton2007}, 
including computer science~\cite{Lawrance2013,Dorodchi2016} and statistics~\cite{Beckman2021}
courses. 
GitHub has played a role in the classroom and for hosting scientific research~\cite{zagalsky2015emergence,feliciano2016student}, 
yet we expect many institutions to be either unwilling or unable to utilize GitHub or other commercial tools~\cite{lerner2002some,van2009adopting,coll2008free}. 
We therefore wish in this work to distinguish between academic and non-academic Penumbra hosts, in order to measure the extent with which academic institutions appear within the Penumbra ecosystem.

The rest of this paper is organized as follows.
In \cref{sec:methods} we describe our materials and methods, how we identify and collect Penumbra projects, how we gather a time-matched sample of GitHub projects, and we describe the subsequent analyses we perform on collected projects and the statistical models we employ.
We report our results in \cref{sec:results} including our analysis of our Penumbra sample and our comparison to our GitHub sample.
\Cref{sec:discussion} concludes with a discussion of our results, limitations of our study, and avenues for future work.

\section{Materials and methods}
\label{sec:methods}

\subsection{Data collection}
\label{subsec:datasets}

We began by identifying various open source software packages that can serve as self-hosted alternatives to GitHub.
These included GitLab Community Edition (CE), Gitea, Gogs, cgit, RhodeCode, and SourceHut.
We limited ourselves to platforms with a web-git interface similar to mainstream centralized platforms like GitHub and GitLab, and so chose to exclude command-line only source code management like GitoLite, as well as more general project management software like Jitsi and Phabricator.
For each software package, we identified a snippet of HTML from each package's web interface that uniquely identifies that software.
Often this was a version string or header, such as \texttt{<meta content="GitLab" property="og:site\_name">}.

We then turned to Shodan~\cite{shodan} to find hosts running instances of each software package.
Shodan maintains a verbose port scan of the entire IPv4 and some of the IPv6 Internet, including response information from each port, such as the HTML returned by each web server.
This port scan is searchable, allowing us to list all web servers open to the public Internet that responded with our unique identifier HTML snippets.
Notably, Shodan scans only include the default web page from each host, so if a web server hosts multiple websites and returns different content depending on the host in the HTTP request, then we will miss all but the front page of the default website.
Therefore, Shodan results should be considered a strict under-count of public instances of these software packages. However, we have no reason to believe that it is a biased sample, as there are trade-offs to dedicated and shared web hosting for organizations of many sizes and purposes.

We narrowed our study to the three software packages with the largest number of public instances: GitLab CE, Gogs, and Gitea.
Searching Shodan, we found 59596 unique hosts.
We wrote a web crawler for each software package, which would attempt to list every repository on each host, and would report when instances were unreachable (11677), had no public repositories (44863), or required login information to view repositories (2101).
We then attempted to clone all public repositories, again logging when a repository failed to clone, sent us a redirect when we tried to clone, or required login information to clone.
For each successfully cloned repository, we checked the first commit hash against the GitHub API, and set aside repositories that matched GitHub content (see \cref{subsec:methods:findingDuplicates}).
We discarded all empty (zero-commit) repositories.
This left us with 45349 repositories from 1558 distinct hosts.

Next, we wanted to collect a sample of GitHub repositories to compare development practices. 
We wanted a sample of a similar number of repositories from a similar date range, to account for trends in software development and other variation over time.
We chose not to control for other repository attributes, like predominant programming language, size of codebase or contributorship, or repository purpose. 
We believe these attributes may be considered factors when developers choose where to host their code, so controlling for them would inappropriately constrain our analysis.
To gather this comparison sample, we drew from GitHub Archive~\cite{grigorik2012github} via their BigQuery interface to find an equal number of ``repository creation'' events from each month a Penumbra repository was created in. 
We attempted to clone each repository, but found that some repositories had since been deleted, renamed, or made private.
To compensate, we oversampled from GitHub Archive for each month by a factor of 1.5.
After data collection and filtering we were left with a time-matched sample of 57914 GitHub repositories.

Lastly, to help identify academic hosts, we used a publicly available list of university domains\footnote{\url{https://github.com/hipo/university-domains-list}}. This is a community-curated list, and so may contain geographic bias, but was the most complete set of university domains we located.

\subsection{Host analysis}
\label{subsec:hostanalysis}

We used geoip lookups\footnote{\url{https://dev.maxmind.com/geoip/geolite2-free-geolocation-data/}} to estimate the geographic distribution of hosts found in our Penumbra scan. 
We also created a simple labelling process to ascertain how many hosts were universities or research labs: Extract all unique emails from commits in each repository, and label each email as academic if the hostname in the email appears in our university domain list. 
If over 50\% of unique email addresses on a host are academic, then the host is labeled as academic. 
This cutoff was established experimentally after viewing the distribution of academic email percentages per host, shown in the inset of \cref{fig:overview}(c). 
Under this cutoff, 15\% of Penumbra hosts (130) were tagged as academic.

\subsection{Repository analysis}
\label{subsec:repoanalysis}

We are interested in diverging software development practices between GitHub and the Penumbra, and so we measured a variety of attributes for each repository.
To analyze the large number of commits in our dataset, we modified git2net~\cite{git2net} and PyDriller~\cite{pydriller} to extract only commit metadata, ignoring the contents of binary ``diff'' blobs for performance.
We measured the number of git branches per repository (later, in \cref{fig:github_comparison_no_time}, we count only remote branches, and ignore \texttt{origin/HEAD}, which is an alias to the default branch), but otherwise concerned ourselves only with content in the main branch, so as to disambiguate measurements like ``number of commits.'' 

From the full commit history of the main branch we gather the total number of commits, the hash and time of each commit, the length in characters of each commit message, and the number of repository contributors denoted by unique author email addresses. (Email addresses are not an ideal proxy for contributors; a single contributor may use multiple email addresses, for example if they have two computers that are configured differently. Unfortunately, git commit data does not disambiguate usernames.
Past work~\cite{tutko2020more,gote2021gambit} has attempted to disambiguate authors based on a combination of their commit names and commit email addresses, but we considered this out of scope for our work. By not applying identity disambiguation to either the Penumbra or GitHub repositories, the use of emails-as-proxy is consistent across both samples. If identity disambiguation would add bias, for example if disambiguation is more successful on formulaic university email addresses found on academic Penumbra hosts than it is on GitHub data, then using emails as identifiers will provide a more consistent view.)
From the current state (head commit of the main branch) of the repository we measure the number of files per repository.
This avoids ambiguity where files may have been renamed, split, or deleted in the commit history. 
We apply \textit{cloc}\footnote{\url{https://github.com/AlDanial/cloc}}, the ``Count Lines of Code'' utility, to identify the top programming language per repository by file counts and by lines of code. %

We also calculate several derived statistics. 
The average \textit{interevent time}, the average number of seconds between commits per repository, serves as a crude indicator of how regularly contributions are made. 
We refine this metric as \textit{burstiness}, a measure of the index of dispersion (or Fano Factor) of commit times in a repository \cite{choudhary2018modeling}.
The index of dispersion is defined as $\sigma^2_w/\mu_w$, or the variance over the mean of events over some time window $w$. 
Previous work defines ``events'' broadly to encompass all GitHub activity, such as commits, issues, and pull requests. 
To consistently compare between platforms, we define ``events'' more narrowly as ``commits per day''. 
Note that while interevent time is only defined for repositories with at least two commits, burstiness is defined as 0 for single-commit repositories.

We infer the age of each repository as the amount of time between the first and most recent commit.
One could compare the start or end dates of repositories using the first and last commit as well, but because we sampled GitHub by finding repositories with the same starting months as our Penumbra repositories, these measurements are less meaningful within the context of our study.

Following \citet{bagrow_teams_2016}, we compute three measures for how work is distributed across members of a team.
The first, \textit{lead workload}, is the fraction of commits performed by the ``lead'' or heaviest contributor to the repository.
Next, a repository is \textit{dominated} if the lead
makes more commits than all other contributors combined (over 50\% of commits). 
Note that all single-contributor repositories are implicitly dominated by that single user, and all two-contributor repositories are dominated unless both contributors have an exactly equal number of commits, so dominance is most meaningful with three or more contributors.
Lastly, we calculate an \textit{effective team size}, estimating what the effective number of team members would be if all members contributed equally.
Effective team size $m$ for a repository with $M$ contributors is defined as $m = 2^h$, where $h = - \sum_{i=1}^{M}{f_i \log_2 f_i}$, and $f_i = w_i / W$ is the fraction of work conducted by contributors $i$.
For example, a team with $M=2$ members who contribute equally ($f_1=f_2$) would also have an effective team size of $m=2$, whereas a duo where one team member contributes 10 times more than the other would have an ``effective'' team size of $m=1.356$. 
Effective team size is functionally equivalent to the Shannon entropy $h$, a popular index of diversity, but is exponentiated so values are reported in numbers of team members as opposed to the units of $h$, which are typically bits or nats.
Since we only consider commits as work (lacking access to more holistic data on bug tracking, project management, and other non-code contributions \cite{casari2021open}), $f_i$ is equal to the fraction of commits in a repository made by a particular contributor. Interpreting the contents of commits to determine the magnitude of each contribution (as in expertise-detection studies like \cite{da2015niche}) would add nuance, but would require building parsers for each programming language in our dataset, and requires assigning a subjective value for different kinds of contributions, and so is out of scope for our study.
Therefore, the effective team size metric improves on a naive count of contributors, which would consider each contributor as equal even when their numbers of contributions differ greatly.

\subsection{Duplication and divergence of repositories}
\label{subsec:methods:findingDuplicates}

It is possible for a repository to be an exact copy or ``mirror'' of another repository and this mirroring may happen across datasets: a Penumbra repository could be mirrored on GitHub, for example.
Quantifying the extent of mirroring is important for determining whether the Penumbra is a novel collection of open source code or if it mostly already captured within, for instance, GitHub.
Likewise, a repository may have been a mirror at one point in the past but subsequent edits have caused one mirror to diverge from the other.

Searching for git commit hashes provides a reliable way to detect duplicate repositories, as hashes are derived from the cumulative repository contents\footnote{Commit hashes include the files changed by the commit, and the hash of the parent commit, referencing a list of changes all the way to the start of the repository.} and, barring intentional attack~\cite{stevens2017first} on older versions, hash collisions are rare.
To determine the novelty of Penumbra repositories, we searched for their commit hashes on GitHub, on Software Heritage (SH), a large-scale archive of open source code~\cite{abramatic2018building} and within the Penumbra sample itself to determine the extent of mirroring within the Penumbra.
Search APIs were used for GitHub and SH, while the Penumbra sample was searched locally.
For each Penumbra repository, we searched for the first hash and, if the repository had more than one commit, the latest hash.
If both hashes are found at least once on GitHub or SH, then we have a complete copy (at the time of data collection). If the first hash is found but not the second, then we know a mirror exists but has since diverged. If nothing is found, it is reasonable to conclude the Penumbra project is novel (i.e., independent of GitHub and SH).

To ensure a clean margin when comparing the Penumbra and GitHub samples, we excluded from our analysis (\cref{subsec:repoanalysis}) any Penumbra repositories that were duplicated on GitHub, even if those duplicates diverged.

\subsection{Statistical models}
\label{subsec:methods:statisticalmodels}

To understand better what features most delineate Penumbra and GitHub projects, we employ two statistical models: logistic regression and a random forest ensemble classifier.
While both can in principle be used to predict whether a project belongs to the Penumbra or not, our goal here is inference: we wish to understand what features are most distinct between the two groups.

For logistic regression, we fitted two models.
Exogenous variables were 
numbers of files, %
contributors, %
commits, %
and branches; %
average commit message length; %
average editors per file; %
average interevent time, in hours; %
lead workload, the proportion of commits made by the heaviest contributor; %
effective team size; %
burstiness, as measured by the index of dispersion;
and, for model 1 only, the top programming language as a categorical variable. %
Given differences in programming language choice in academic and industry \cite{rabai2015programming}, we wish to investigate any differences when comparing Penumbra and GitHub projects (see also \cref{subsec:datasets,subsec:languages}).
There is a long tail of uncommon languages that prevents convergence when fitting model 1, so we processed the categorical variable by combining Bourne and Bourne Again languages and grouping languages that appeared in fewer than 1000 unique repositories into an ``other'' category before dummy coding. 
JavaScript, the most common language, was chosen as the baseline category.
Missing values were present, due primarily to a missing top language categorization and/or an undefined average interevent time. 
Empty or mostly empty repositories, as well as repositories with a single commit, will cause these issues, so we performed listwise deletion on the original data, removing repositories from our analysis when any fields were missing.
After processing, we were left with 67,893 repositories (47.26\% Penumbra).
Logistic models were fitted using Newton-Raphson and odds $e^\beta$ and 95\% CI on odds were reported.

For the random forest model, feature importances were used to infer which features were most used by the model to distinguish between the two groups. 
We used the same data as logistic regression model 2, randomly divided into 90\% training, 10\% validation subsets.
We fit an ensemble of 1000 trees to the training data using 
default hyperparameters; random forests were fit using scikit-learn v0.24.2.
Model performance was assessed using an ROC curve on the validation set.
Feature importances were measured with permutation importance, a computationally-expensive measure of importance but one that is not biased in favor of features with many unique values~\cite{strobl2007biasImportance}.
Permutation importance was computed by measuring the fitted model's accuracy on the validation set; then, the values of a given feature were permuted uniformly at random between validation observations and validation accuracy was recomputed. 
The more accuracy drops, the more important that feature was.
Permutations were repeated 100 times per feature and the average drop in accuracy was reported. 
Note that permutation importance may be negative for marginally important features and that importance is only useful as a relative quantity for ranking features within a single trained (ensemble) model.

\section{Results}
\label{sec:results}

We sampled the Penumbra of the open-source ecosystem: Public repositories on public hosts independent from large centralized platforms. Our objective is to compare the Penumbra to GitHub, the largest centralized platform, to better understand the representativeness of GitHub as a sample of the open-source ecosystem and how the choice of platforms might influence online collaborations.
In \cref{subsec:overview} we begin with an overview of the Penumbra's geographic distribution and the scale of hosts.
In \cref{subsec:editing} we analyze the collaboration patterns and temporal features of Penumbra and GitHub repositories.
\Cref{subsec:languages} examines the programming language domains of Penumbra and GitHub projects while \cref{subsec:acnonachosts} further investigates differences between academic and non-academic Penumbra repositories.
Statistical models in \cref{subsec:statisticalmodels} summarize the combined similarities and differences between Penumbra and GitHub repositories.
Finally, in \cref{subsec:novelty} we investigate the novelty of our Penumbra sample, how many Penumbra repositories are duplicates and whether Penumbra repositories also exist on GitHub and within the Software Heritage~\cite{abramatic2018building} archive.

\subsection{An overview of the Penumbra sample}
\label{subsec:overview}

\begin{figure}
    \centering
    \includegraphics[width=0.98\linewidth]{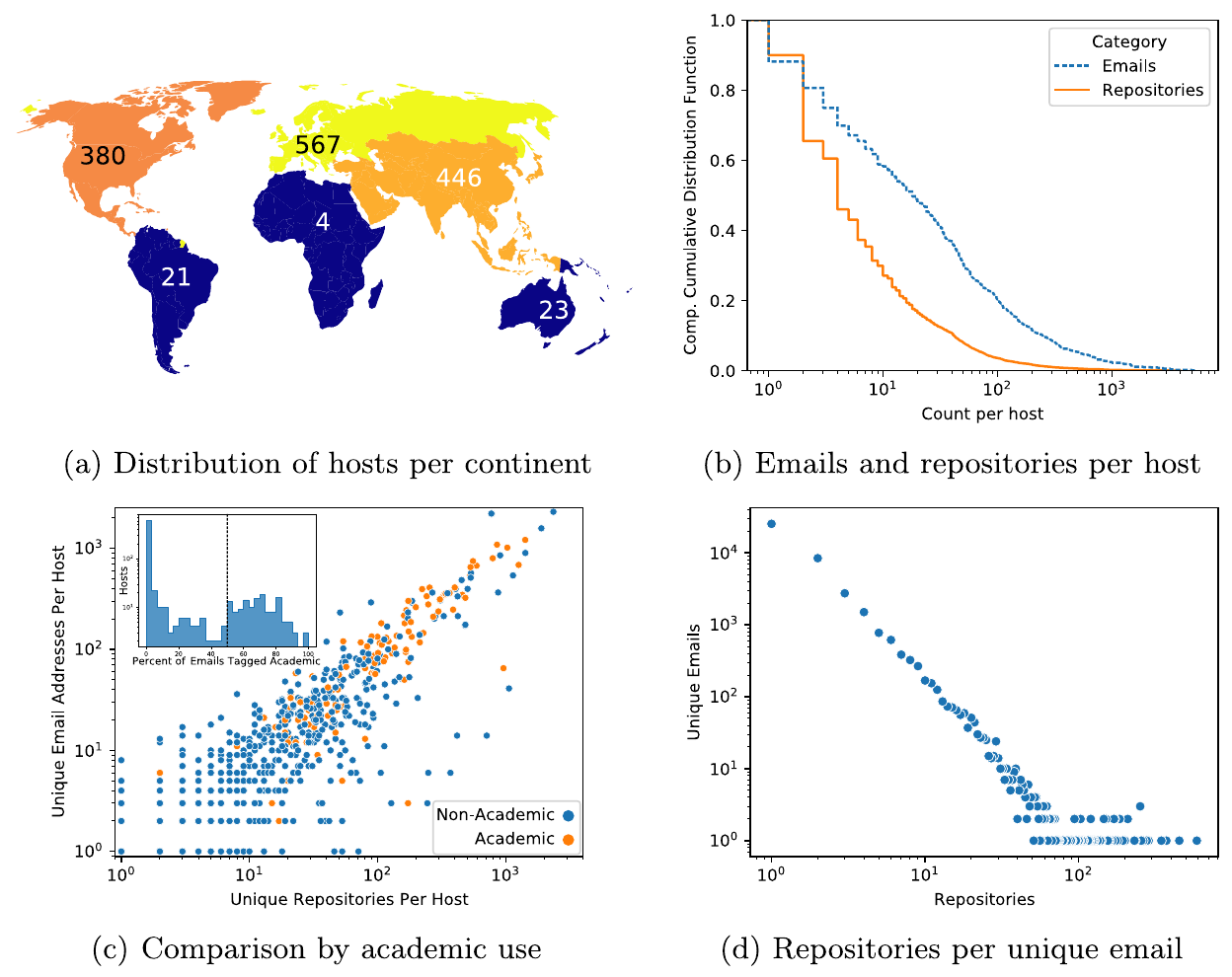}
    \caption{
    \textbf{The penumbra of open source.}
    (a) 
    Geographic distribution of hosts and unique email addresses (in parentheses) in our Penumbra sample.
    (b) 
    Distributions of emails per host and repositories per host.
    (c) Distribution of unique emails per repositories.
    (d) Correlation between repositories and emails per host.
    We see that the number of repositories and email addresses generally correlate, with some outlying hosts with many more repositories than emails. 
    Academic hosts follow the same general trend, except that they tend to be larger than many non-academic hosts.
    (inset) Hosts are classified as ``academic" if over 50 percent of their email addresses end in \texttt{.edu} or come from a manually identified academic domain.
    }
    \label{fig:overview}
\end{figure}

Our Penumbra sample consists of 1558 distinct hosts from all six inhabited continents and 45349 non-empty repositories with no matching commits on GitHub (\cref{subsec:methods:findingDuplicates}; we explore overlap with GitHub in \cref{subsec:novelty}).
This geographic distribution, illustrated in \cref{fig:overview} and described numerically in \cref{table:repos_per_continent}, shows that the Penumbra is predominantly active in Europe, North America, and Asia by raw number of hosts and repositories. However, Oceania has the second most repositories per capita, and the highest percentage of academic emails in commits from repositories cloned from those hosts (\cref{table:repos_per_continent}). Overall, the geographic spread of the Penumbra is similar to GitHub's self-reported distribution of users~\cite{2020octoverse}, but with a stronger European emphasis and even less Southern Hemisphere representation. 

\begin{table}%
\centering
\caption{Geographic split of our Penumbra (PN) and GitHub (GH)~\cite{2020octoverse} samples.}
\footnotesize
\makebox[\textwidth][c]{%
\begin{tabular}{rrrrrrr}\toprule
Region & \% Hosts & \% PN users & \% GH users~\cite{2020octoverse} & PN repositories (per capita) & \% Unique emails from academic domains \\ \midrule
EU &  39.35 & 73.47 & 26.8 & 83612 ($1.52 \times 10^{-4}$) & 39.20 \\
NA &  26.37 & 15.81 & 34 & 51245 ($2.97 \times 10^{-4}$) & 41.22 \\
AS &  30.95 &  7.38 & 30.7 & 21765 ($1.55 \times 10^{-5}$) & 1.21 \\
SA &   1.46 &  1.36 & 4.9 & 2776  ($1.64 \times 10^{-5}$) & 12.41 \\
OC &   1.60 &  1.93 & 1.7 & 3347  ($2.58 \times 10^{-4}$) & 65.24 \\
AF &   0.28 &  0.04 & 2 & 215   ($9.44 \times 10^{-7}$) & 0.00  \\ \bottomrule
\end{tabular}
\label{table:repos_per_continent}
}
\end{table}

We find a strong academic presence in the Penumbra: on 15\% of hosts, more than half of email addresses found in commits come from academic domains (see also \cref{subsec:acnonachosts}).
These academic hosts make up many of the larger hosts, but represent a minority of all Penumbra repositories (37\% of non-GitHub-mirrors). We plotted the ``size" of each host in terms of unique emails and repositories, as well as its academic status, in \cref{fig:overview}(c). We find that while academic hosts tend not to be ``small", they do not dominate the Penumbra in terms of user or repository count, refuting the hypothesis that most Penumbra activity is academic.

We are also interested in how distinct hosts are: How many repositories do users work on, and are those repositories all on a single host, or do users contribute to code on multiple hosts?
To investigate, we first plot the number of unique email addresses per host in \cref{fig:overview}(b), then count the number of email addresses that appear on multiple hosts.
Critically, users may set a different email address on different hosts (or even unique emails per-repository, or per-commit, although this would be tedious and unlikely), so using email addresses as a proxy for ``shared users" offers only a lower-bound on collaboration.
We find that 91.7\% of email addresses in our dataset occur on only one host, leaving 3435 email addresses present on 2-4 hosts.
Fifteen addresses appear on 5-74 hosts, but all appear to be illegitimate, such as ``you@example.com", emails without domain suffixes like ``admin" or ``root@localhost", and a few automated systems like ``anonymous@overleaf.com".
We find 61 email addresses on hosts in two or more countries (after removing fake email addresses by the aforementioned criteria), and 33 on multiple continents (after the same filtering).

We did not repeat this analysis on our GitHub sample, because the dataset is too different for such a comparison to be meaningful. All GitHub repositories are on a single ``host", so there is no analogue to ``multi-host email addresses". We considered comparing distributions of ``repositories committed to by each email", but ruled this out because of our data collection methodology. For each Penumbra host, we have data on every commit in every public repository, giving us a complete view of each user's contributions. For GitHub however, we have a small sample of repositories from the entire platform, so we are likely to miss repositories that each GitHub user contributed to.

\subsection{Collaboration patterns and temporal features}
\label{subsec:editing}

\begin{figure}[t!]
    \centering
    \includegraphics[width=0.98\linewidth]{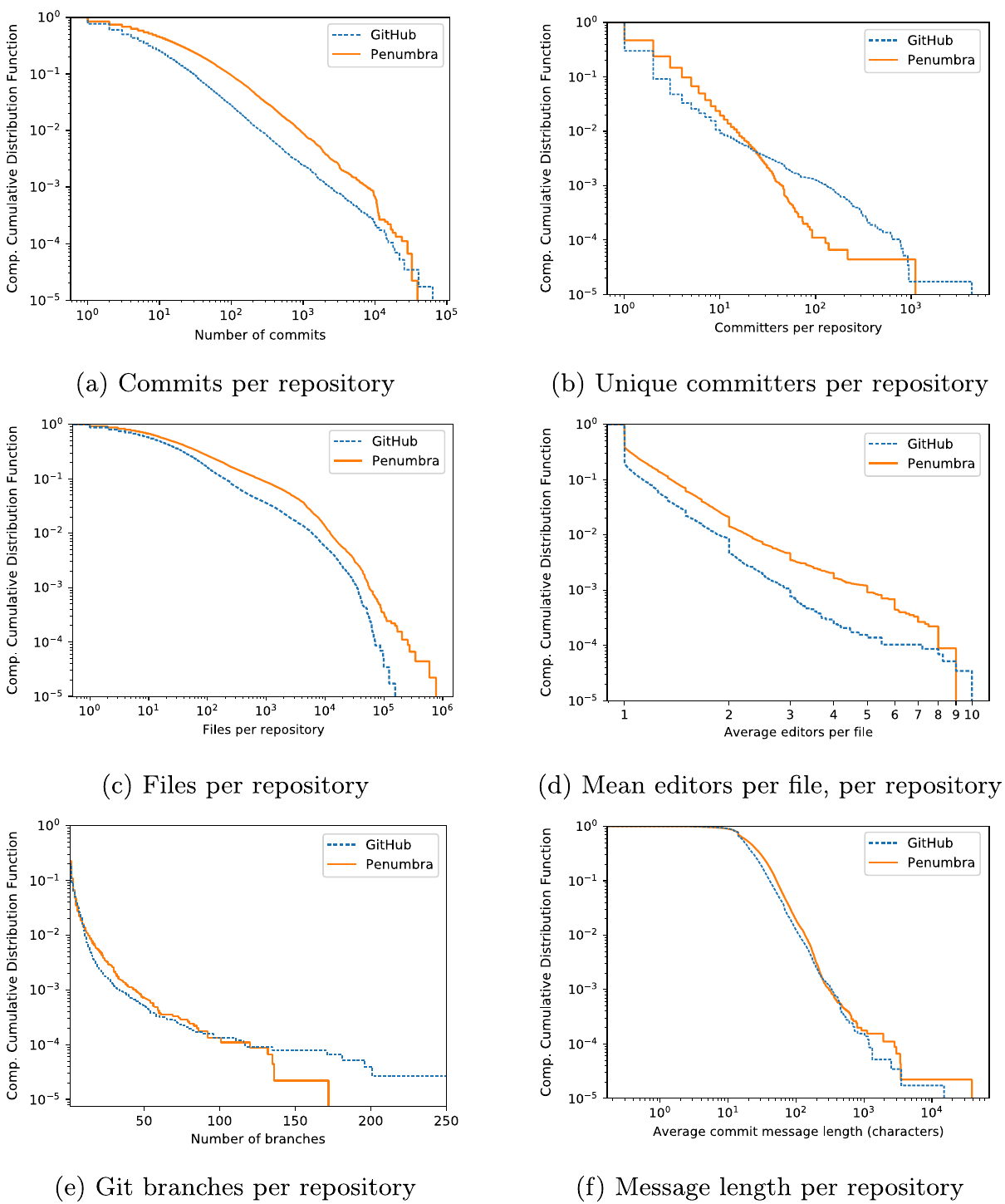}
    \begin{subfigure}[b]{0\textwidth}
         \phantomcaption
         \label{fig:compare_commits}
    \end{subfigure}
    \begin{subfigure}[b]{0\textwidth}
         \phantomcaption
         \label{fig:compare_contributors}
    \end{subfigure}
    \begin{subfigure}[b]{0\textwidth}
         \phantomcaption
         \label{fig:compare_files}
    \end{subfigure}
    \begin{subfigure}[b]{0\textwidth}
         \phantomcaption
         \label{fig:compare_editor_density}
    \end{subfigure}
    \begin{subfigure}[b]{0\textwidth}
         \phantomcaption
         \label{fig:compare_branches}
    \end{subfigure}
    \begin{subfigure}[b]{0\textwidth}
         \phantomcaption
         \label{fig:compare_message_lengths}
    \end{subfigure}
    \caption{%
    \textbf{Editing and collaborating in the Penumbra and GitHub.}
    Comparison of GitHub and Penumbra samples on a variety of metrics. 
    Unique users for all plots are determined by unique email addresses in commit data. File counts are taken at the HEAD commit of the main branch. Editor overlap is defined as average number of unique contributors that have edited each file. Panel (e) excludes two GitHub repositories with 500 and 1300 branches, to make trend comparison easier.
    }
    \label{fig:github_comparison_no_time}
\end{figure}

We compare software development and collaboration patterns between our Penumbra sample and a GitHub sample of equivalent size and time period (\cref{fig:github_comparison_no_time,table:statistics_compared}). We examine commits per repository, unique emails per repository (as a proxy for unique contributors), files per repository, average editors per file, branches per repository, and commit message length. While mean behavior was similar in both repository samples, diverging tail distributions show that Penumbra repositories usually have more commits, more files, fewer emails, and more editors per file.

\begin{table}
\caption{Comparison of Penumbra and GitHub datasets} %
\footnotesize
\makebox[\textwidth][c]{%
\begin{tabular}{llllllllll}
\toprule
\multicolumn{2}{c}{} & \multicolumn{3}{l}{Penumbra} & \multicolumn{3}{l}{GitHub} & \multicolumn{2}{l}{KS 2-Sample} \\
Statistic & Fig. & Mean & Median & CI & Mean & Median & CI & KS S & KS P \\ \midrule
Files & \ref{fig:github_comparison_no_time}(c) & 244.47 & 12 & [1,859] & 156.07 & 9 & [1,264] & 0.07 & $<$ 0.001 \\
Committers & \ref{fig:github_comparison_no_time}(b) & 2.39 & 1 & [1,6] & 2.08 & 1 & [1,3] & 0.17 & $<$ 0.001 \\
Message Lengths & \ref{fig:github_comparison_no_time}(f) & 29.24 & 20.80 & [7.00,67.33] & 24.23 & 17.60 & [7.42,56.00] & 0.13 & $<$ 0.001 \\
Editor Density & \ref{fig:github_comparison_no_time}(d) & 1.12 & 1.00 & [1.00,1.60] & 1.05 & 1.00 & [1.00,1.30] & 0.20 & $<$ 0.001 \\
Burstiness & \ref{fig:github_comparison_time}(d) & 4.86 & 2.88 & [0.50,14.51] & 3.68 & 2.15 & [0.17,11.24] & 0.13 & $<$ 0.001 \\
Commits & \ref{fig:github_comparison_no_time}(a) & 67.12 & 8 & [1,194] & 25.27 & 4 & [1,57] & 0.20 & $<$ 0.001 \\
Branches & \ref{fig:github_comparison_no_time}(e) & 1.74 & 1 & [1,4] & 1.67 & 1 & [1,5] & 0.03 & $<$ 0.001 \\
Age (hours) & \ref{fig:github_comparison_time}(a) & 5528 & 883 & [0.1,25556] & 2669 & 73 & [0.03,16194] & 0.26 & $<$ 0.001 \\
Age / Commits & \ref{fig:github_comparison_time}(b) & 283 & 39 & [0.02,1261] & 193 & 9 & [0.01,944] & 0.19 & $<$ 0.001 \\
Avg. Interevent & \ref{fig:github_comparison_time}(c) & 375 & 43 & [0.05,1547] & 257 & 11 & [0.02,1130] & 0.19 & $<$ 0.001 \\
Team Size & \ref{fig:github_comparison_time}(e) & 1.71 & 1.00 & [1.00,3.92] & 1.42 & 1.00 & [1.00,2.67] & 0.17 & $<$ 0.001 \\
\bottomrule
\end{tabular}}
\begin{flushleft}
\footnotesize
\medskip \par
Legend: Mean, median, and 5\textsuperscript{th} and 95\textsuperscript{th} percentile values from the Penumbra and GitHub samples for each statistic. \textbf{KS S} and \textbf{KS P} represent the Kolmogorov-Smirnov two-sample statistic, and its corresponding p-value. 
\end{flushleft}
\label{table:statistics_compared}
\end{table}

\begin{figure}[t!]%
    \centering
    \includegraphics[width=0.98\linewidth]{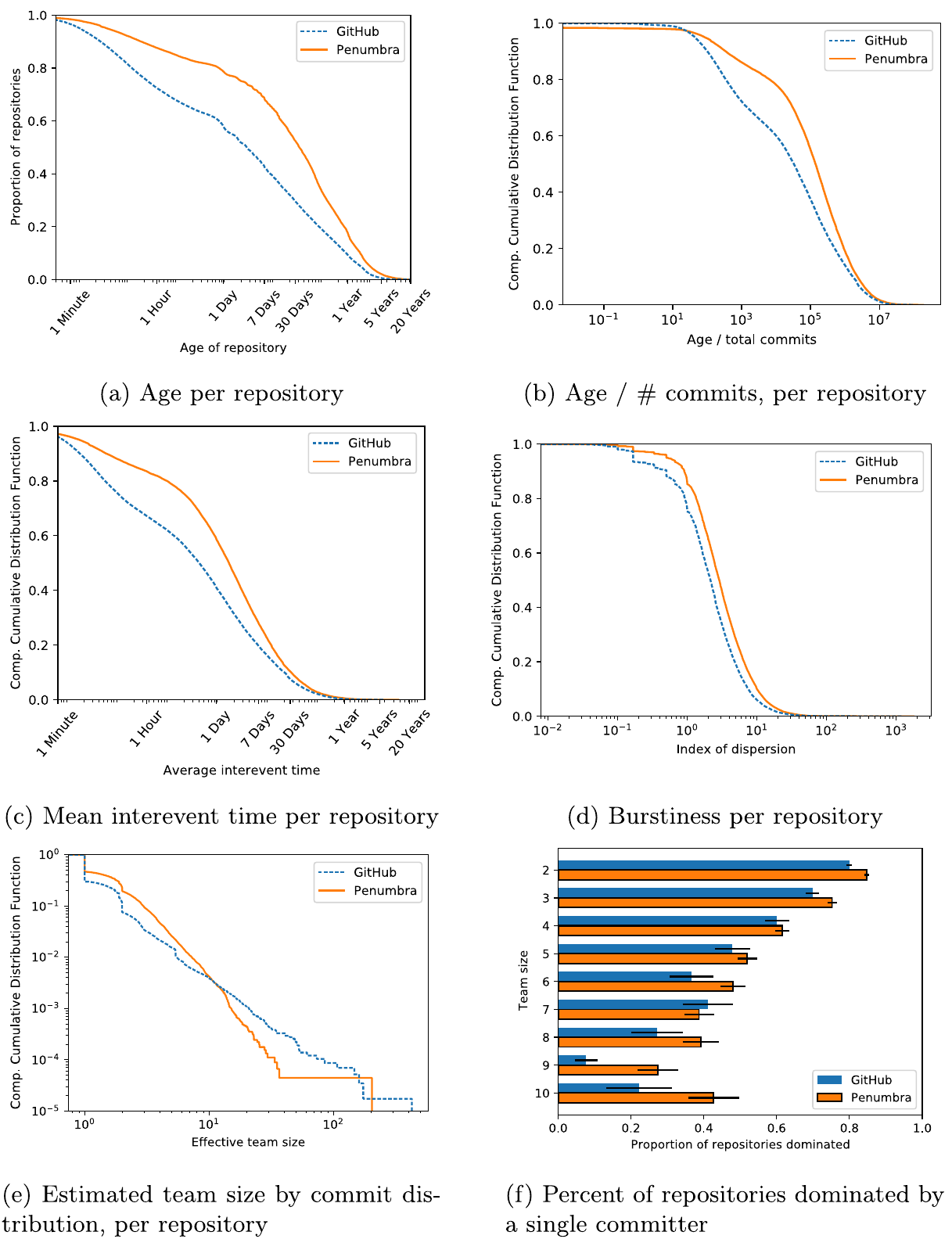}
    \begin{subfigure}[b]{0\textwidth}
         \phantomcaption
         \label{fig:compare_age}
    \end{subfigure}
    \begin{subfigure}[b]{0\textwidth}
         \phantomcaption
         \label{fig:compare_age_over_comits}
    \end{subfigure}
    \begin{subfigure}[b]{0\textwidth}
         \phantomcaption
         \label{fig:compare_avg_interevent}
    \end{subfigure}
    \begin{subfigure}[b]{0\textwidth}
         \phantomcaption
         \label{fig:compare_burstiness}
    \end{subfigure}
    \begin{subfigure}[b]{0\textwidth}
         \phantomcaption
         \label{fig:compare_teamsize}
    \end{subfigure}
    \begin{subfigure}[b]{0\textwidth}
         \phantomcaption
         \label{fig:compare_dominance}
    \end{subfigure}
    \caption{
    \textbf{Temporal characteristics of collaboration in the Penumbra and GitHub.}
    Comparison of GitHub and Penumbra samples on a variety of temporal metrics. The age of repository is given by the time between the first and latest commit. Panels (b-d) look at the heterogeneity of time between events. We first compare the distribution of mean interevent time to the distribution of ratios of age to number of commits, then show the distribution of index of dispersion per repository. Panels (e-f) compare how collaborative repositories actually are, or whether they are dominated by a single committer.
    }
    \label{fig:github_comparison_time}
\end{figure}

One might hypothesize that with more files and fewer editors the Penumbra would have stronger ``partitioning", with each editor working on a different subset of files. However, our last three metrics suggest that the Penumbra has more collaborative tendencies: while Penumbra repositories are larger (in terms of files), with smaller teams (in terms of editors), there are on average \textit{more} contributors working on the same files or parts of a project. To deepen our understanding of this collaborative behavior, we also estimated the ``effective team size" for each repository by the fraction of commits made by each editor. This distinguishes consistent contributors from editors with only a handful of commits, such as ``drive-by committers" that make one pull request, improving upon a naive count of unique emails. These estimates show that while there are more GitHub repositories with one active contributor, and more enormous projects with over 50 team members, the Penumbra has more repositories with between 2 and 50 team members. However, for all team sizes between 2 and 10, we find that more penumbra repositories are ``dominated" by a single contributor, see \cref{fig:github_comparison_time}(f), meaning that their top contributor has made over 50\% of all commits.

We also compare temporal aspects of Penumbra and GitHub repositories (\cref{fig:github_comparison_time}). 
Penumbra repositories are shown to be generally older in terms of ``time between the first and most recent commit" in \cref{fig:github_comparison_time}(a), have more commits in \cref{fig:github_comparison_time}(b), but are also shown to have a longer time between commits measured both as interevent time in \cref{fig:github_comparison_time}(c), and as burstiness in \cref{fig:github_comparison_time}(d). This means that while Penumbra repositories are maintained for longer (or conversely, there are many short-lived repositories on GitHub that receive no updates), they are maintained inconsistently or in a bursty pattern, receiving updates after long periods of absence. And while both GitHub and Penumbra repositories tend to be bursty, a larger portion of Penumbra repositories exhibit burstiness as indicated by an index of dispersion above 1.

\subsection{Language domains}
\label{subsec:languages}

\begin{figure}[t]%
    \centering
    \includegraphics[width=0.98\linewidth]{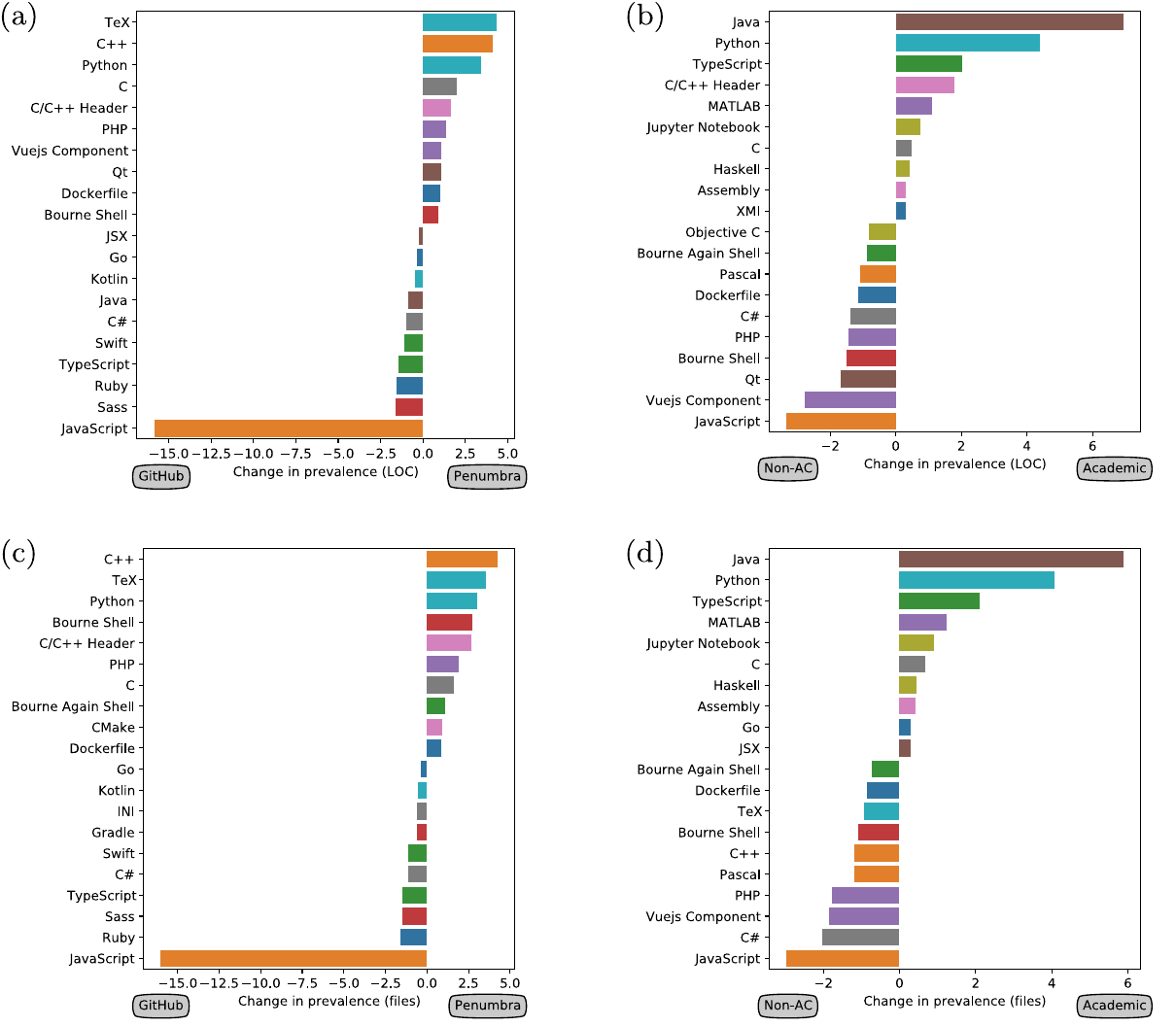}
    \caption{
    \textbf{Dominant language domains in the Penumbra and GitHub}
    Comparison of language popularity, measured by lines of code (LOC) in panels (a-b) and by file count in panels (c-d). We count the top languages of each repository by the specified metric, normalize the results to a percentage of independent or GitHub repositories, then subtract the percentages. Therefore a language with a value of $-0.05$, for example, is the top language on $5\%$ more GitHub repositories than Penumbra repositories, while a positive value indicates $5\%$ more Penumbra repositories than GitHub repositories.
    }
    \label{fig:lang}
\end{figure}

Most of our analysis has focused on repository metadata (commits and files), rather than the content of the repositories. This is because more in-depth content comparison, such as the dependencies used, or functions written within a repository's code, would vary widely between languages and require complex per-language parsing. 
However, we have classified language prevalence across the Penumbra and GitHub by lines of code and file count per repository in \cref{fig:lang}(left column). 
We find that the Penumbra emphasizes academic languages (TeX) and older languages (C, C++, PHP, Python), while GitHub represents more web development (JavaScript, TypeScript, Ruby), and mobile app development (Swift, Kotlin, Java). We additionally compare repositories within the Penumbra that come from academic hosts ($>$ 50\% emails come from academic domains; see Methods) and non-academic hosts, using the same lines of code and file count metrics in \cref{fig:lang}(right column). 
Academic hosts unsurprisingly contain more languages used in research (Python, MATLAB, and Jupyter notebooks), and languages used in teaching (Haskell, assembly, C). Despite Java's prevalence in enterprise and mobile app development, and JavaScript's use in web development, academic hosts also represent more Java and Typescript development.
By contrast, non-academic hosts contain more desktop or mobile app development (Objective C, C\#, QT), web development (JavaScript, PHP), shell scripts and docker files, and, surprisingly, Pascal.

\subsection{Academic and non-academic hosts}
\label{subsec:acnonachosts}
    
\begin{figure}%
    \centering
    \includegraphics[width=0.98\linewidth]{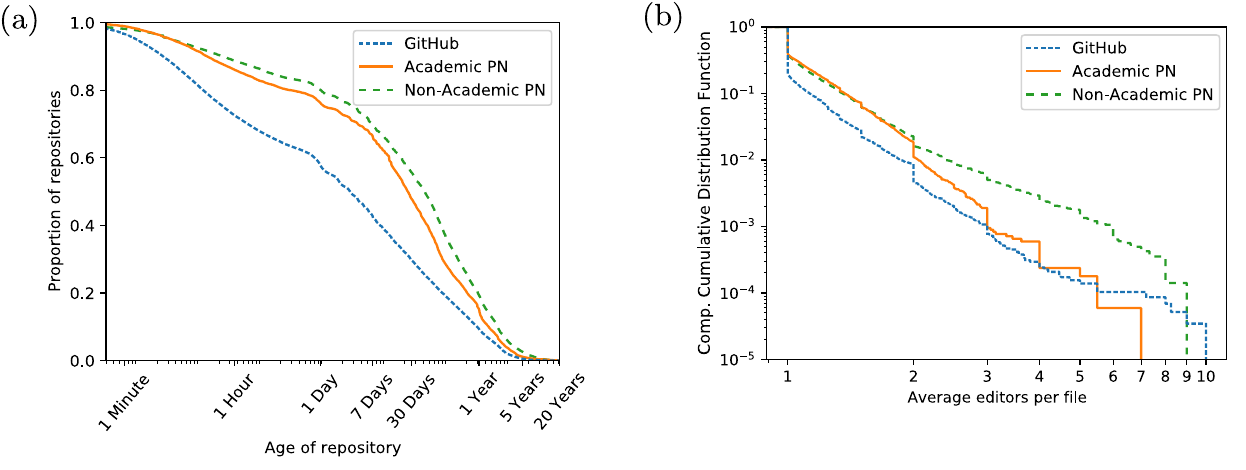}
    \caption{
    \textbf{Comparing academic and non-academic Penumbra repositories to GitHub}
Fifteen percent of Penumbra hosts are ``academic" under our definition, representing 37\% of all Penumbra repositories.
We find that academic repositories are maintained for about as long as non-academic Penumbra repositories, so academic development practices do not drive the divergence from GitHub development patterns that we observe.
Academic repositories have fewer editors per file than non-academic Penumbra repositories, however, more closely matching development practices seen on GitHub.
This refutes the hypothesis that the Penumbra differs widely from GitHub primarily due to academic influence.
}
    \label{fig:ac-non-ac-breakdown}
\end{figure}

Academic hosts account for over 15\% of hosts and 37\% of repositories in the Penumbra, so one might hypothesize that academic software development has a striking effect on the differences between GitHub and the Penumbra. 
To investigate this, \cref{fig:ac-non-ac-breakdown} redraws \cref{fig:compare_age,fig:compare_editor_density} with academic and non-academic Penumbra repositories distinguished. 
We find that the academic repositories are maintained for about the same length of time as their non-academic counterparts, and that academic repositories have fewer editors per file than non-academic development. 
In fact, academic repositories more closely match GitHub repositories in terms of editors per file. 
Therefore, we find that academic software development does not drive the majority of the differences between GitHub and the Penumbra.

\subsection{Statistical models}
\label{subsec:statisticalmodels}

To understand holistically how these different features delineate the two data sources, we perform combined statistical modeling.
First, we performed logistic regression (\cref{tab:logistics}) on the outcome variable of GitHub vs.\ Penumbra, see \cref{subsec:methods:statisticalmodels} for details. 
We fit two models, one containing the primary programming language as a feature and the other not. Examining the odds $e^\beta$ for each variable, we can determine which variables, with other variables held constant, most clearly distinguish GitHub and Penumbra repositories.
The strongest non-language separators are average editors per file, lead workload, and the number of contributors. 
The strongest language separators are TeX, C/C++ Headers, and C++.
The odds on these variables underscore our existing results: Penumbra projects have more editors per file and less workload placed upon the lead contributor.
Likewise, the odds on TeX and C/C++ code make it more likely for Penumbra projects to be focused on academic and scientific problems.

\newcommand{\pvalsmall}[0]{$< 0.001$}

\begin{table}[t]
    \centering
    \caption{Logistic regression models for GitHub vs. Penumbra outcome. 
    \label{tab:logistics}
    }
\footnotesize
\makebox[\textwidth][c]{%
\begin{tabular}{lllllll}
\toprule
{} & \multicolumn{3}{l}{Model 1} & \multicolumn{3}{l}{Model 2} \\
{} & $e^\beta$ &   $p$ &               CI & $e^\beta$ &   $p$ &             CI \\
\midrule
Constant                          &     0.188 & \pvalsmall{} &    [0.156,0.225] &     0.435 & \pvalsmall{} &  [0.364,0.520] \\
\textbf{Language} (vs.\ JavaScript)         &           &       &                  &           &       &                \\
\hspace{3mm} Bourne (Again) Shell &     3.478 & \pvalsmall{} &    [3.162,3.826] &           &       &                \\
\hspace{3mm} C                    &     4.065 & \pvalsmall{} &    [3.671,4.502] &           &       &                \\
\hspace{3mm} C\#                  &     1.589 & \pvalsmall{} &    [1.444,1.750] &           &       &                \\
\hspace{3mm} C++                  &     5.636 & \pvalsmall{} &    [5.184,6.127] &           &       &                \\
\hspace{3mm} C/C++ Header         &     5.829 & \pvalsmall{} &    [5.103,6.657] &           &       &                \\
\hspace{3mm} Java                 &     2.192 & \pvalsmall{} &    [2.070,2.321] &           &       &                \\
\hspace{3mm} Jupyter Notebook     &     2.722 & \pvalsmall{} &    [2.459,3.012] &           &       &                \\
\hspace{3mm} \textit{OTHER}       &     2.124 & \pvalsmall{} &    [2.023,2.230] &           &       &                \\
\hspace{3mm} PHP                  &     2.524 & \pvalsmall{} &    [2.323,2.743] &           &       &                \\
\hspace{3mm} Python               &     2.804 & \pvalsmall{} &    [2.651,2.965] &           &       &                \\
\hspace{3mm} TeX                  &    30.641 & \pvalsmall{} &  [25.348,37.040] &           &       &                \\
\hspace{3mm} TypeScript           &     1.078 & 0.187 &    [0.964,1.205] &           &       &                \\
\hspace{3mm} Vuejs Component      &     4.940 & \pvalsmall{} &    [4.331,5.635] &           &       &                \\
Files                             &     1.000 & \pvalsmall{} &    [1.000,1.000] &     1.000 & \pvalsmall{} &  [1.000,1.000] \\
Commits                           &     1.001 & \pvalsmall{} &    [1.001,1.001] &     1.001 & \pvalsmall{} &  [1.001,1.001] \\
Average editors per file          &     3.337 & \pvalsmall{} &    [3.002,3.709] &     3.328 & \pvalsmall{} &  [3.002,3.689] \\
Average message length            &     1.002 & \pvalsmall{} &    [1.001,1.002] &     1.002 & \pvalsmall{} &  [1.001,1.003] \\
Burstiness                        &     1.059 & \pvalsmall{} &    [1.055,1.063] &     1.058 & \pvalsmall{} &  [1.053,1.062] \\
Average interevent time [h]       &     1.000 & \pvalsmall{} &    [1.000,1.000] &     1.000 & \pvalsmall{} &  [1.000,1.000] \\
Branches                          &     0.971 & \pvalsmall{} &    [0.965,0.976] &     0.961 & \pvalsmall{} &  [0.955,0.966] \\
Lead workload                     &     0.461 & \pvalsmall{} &    [0.407,0.522] &     0.433 & \pvalsmall{} &  [0.384,0.489] \\
Committers                        &     0.945 & \pvalsmall{} &    [0.936,0.955] &     0.940 & \pvalsmall{} &  [0.931,0.950] \\
Effective team size               &     1.010 & 0.516 &    [0.980,1.040] &     1.016 & 0.315 &  [0.985,1.047] \\
-2LL                              & 84478.844 &       &                  & 89788.631 &       &                \\
Pseudo-R2                         &     0.100 &       &                  &     0.044 &       &                \\
\bottomrule
\end{tabular}
}
\end{table}

\begin{figure}[t]
\centering
\includegraphics[width=0.8\textwidth]{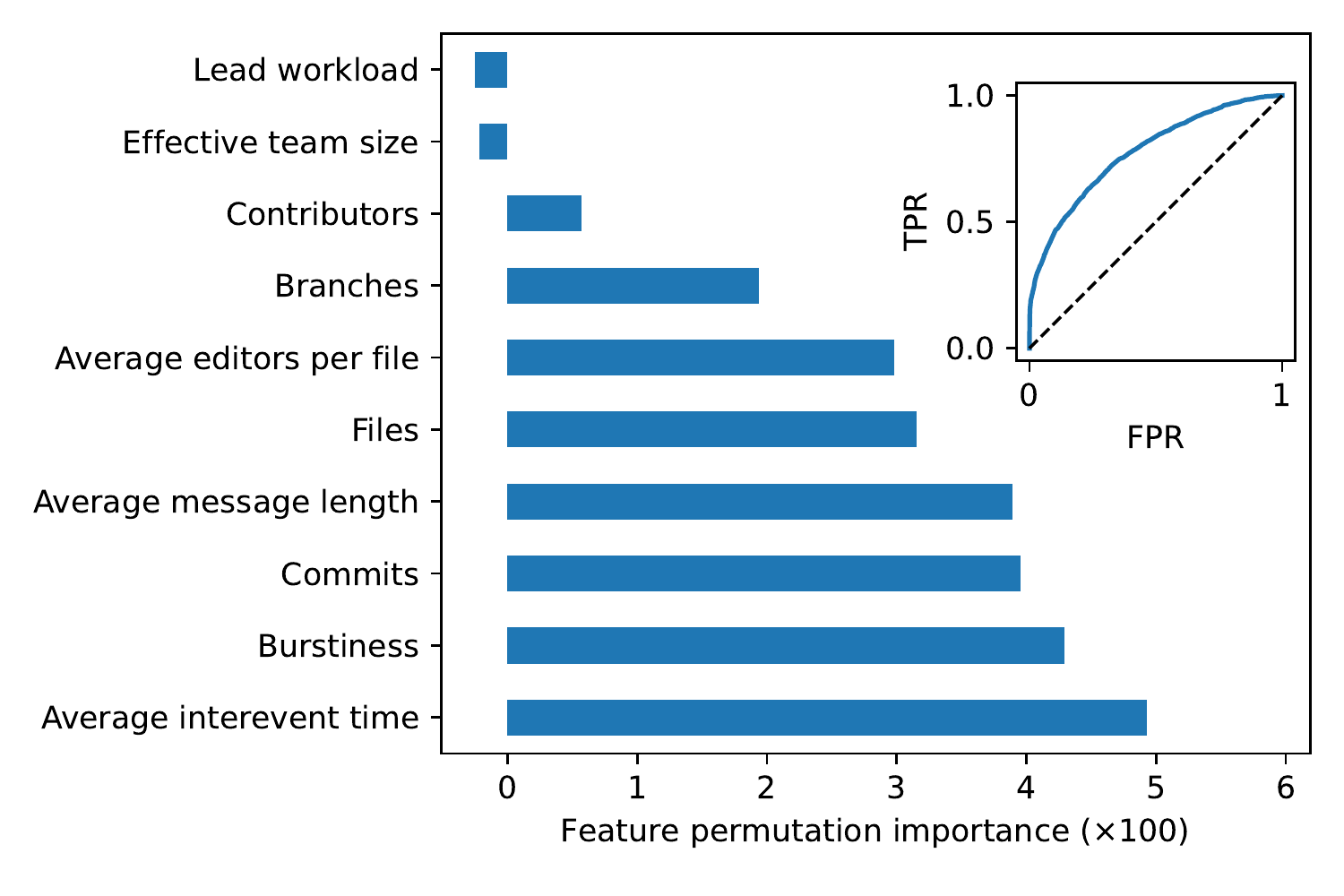}
\caption{\textbf{Random forest model to delineate Penumbra and GitHub samples.}
Feature permutation importance (\cref{subsec:methods:statisticalmodels}) once nonlinear random forest regressions were trained to predict whether a project was on GitHub or in the Penumbra. 
The predictive performance is shown in the inset using an ROC curve of true positive rate (TPR) and false positive rate (FPR).
\label{fig:feature-importance-random-forest}
}
\end{figure}

Supplementing our logistic models we also used nonlinear random forest regressions trained to predict whether a project was in GitHub or the 
Penumbra.
While trained models can be used as predictive classifiers, our goal is to interpret which model features are used to make those predictions, so we report in \cref{fig:feature-importance-random-forest} the top-ten feature importances (\cref{subsec:methods:statisticalmodels}) in our model.
Here we find some differences and similarities with the (linear) logistic regression results. 
Both average editors per file and number of contributors were important, but the random forest found that lead workload was not particularly important. However, the most important features for the random forests were average interevent time, burstiness, and number of commits. (All three were also significant in the logistic regression models.)
The overall predictive performance of the random forest is reasonable (\cref{fig:feature-importance-random-forest} inset).
Taken together, the random forest is especially able to separate the two classes of projects based on time dynamics.

\subsection{Novelty of the Penumbra sample}
\label{subsec:novelty}

\begin{figure}[h!]
    \centering
    \includegraphics[width=0.8\linewidth]{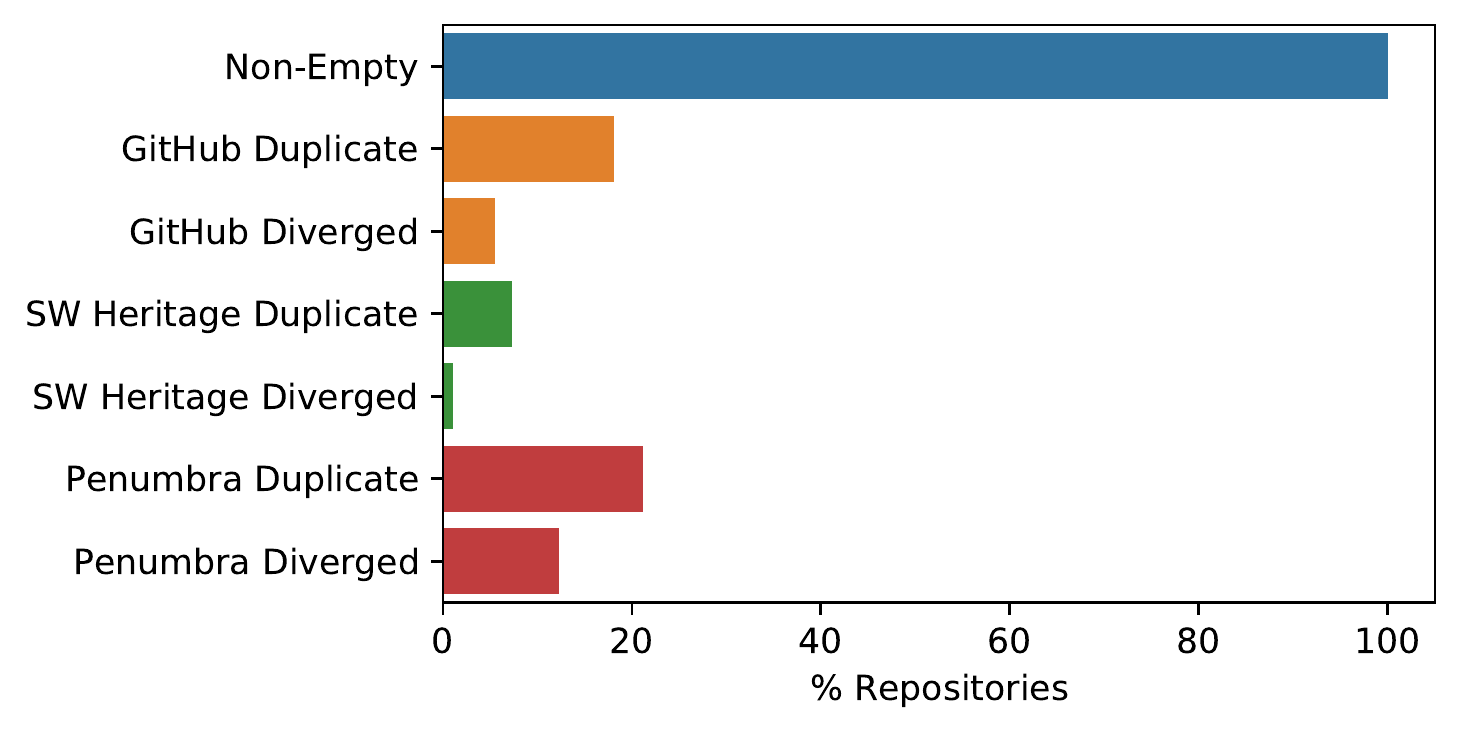}
    \caption{\textbf{The Penumbra's intersection with other datasets.}
    Of the 55343 discovered, non-empty repositories,
    18\% have a first commit hash that can also be found on GitHub (GitHub Duplicate), but 30\% of those repositories diverge and contain code not found on GitHub (GitHub Diverged). Likewise, 7\% of Penumbra repositories have a first commit archived by Software Heritage~\cite{abramatic2018building}, and 14\% of those contain code not archived by Software Heritage. Finally, 21\% of Penumbra repositories share a commit with one or more other Penumbra repositories, and of these, 58\% have unique final commits.
    }
    \label{fig:mirrors}
\end{figure}

How novel are the repositories we have discovered in the Penumbra? 
It may be that many Penumbra repositories are ``mirrored'' on GitHub, in which case the collected Penumbra sample would not constitute especially novel data. 
In contrast, if few repositories appear on GitHub, then we can safely conclude that the Penumbra is a novel collection of open source code.
To test the extent that the Penumbra is independent of GitHub, we checked the first commit hash of each Penumbra repository against the GitHub Search API (\cref{subsec:methods:findingDuplicates}).
We found 9994 such repositories (\cref{fig:mirrors}) and conclude that the majority of Penumbra repositories are novel. 
We excluded these overlapping repositories from our comparisons between the Penumbra and GitHub.
However, such repositories may not represent true duplicates, but instead ``forks'', where developers clone software from GitHub to the Penumbra and then make local changes, or vice-versa, leading to diverging code.
To disambiguate, we checked the \textit{last} commit hash from each of the 9994 overlapping repositories against the GitHub API, and found 3056 diverging commits, as illustrated in \cref{fig:mirrors}.
In other words, 30\% of Penumbra repositories with a first commit on GitHub also contain code \textit{not} found on GitHub.
While we still excluded these repositories to ensure a wide margin between the samples, in fact, the differences in these repositories further underscore the novelty of the Penumbra data.

We also compared our repositories against Software Heritage~\cite{abramatic2018building}, an archive of open source software development.
While Software Heritage is not a hosting platform like GitHub, it represents a potentially similar dataset to our own.
Applying the same methodology as for GitHub mirror detection, we found that 4053 repositories (9\% of our non-empty Penumbra sample) had a matching first commit hash archived on Software Heritage, and that of these, 564 repositories (14\% of overlapping first commits) contained code \textit{not} archived by Software Heritage.
Since Software Heritage is an archive, rather than a software development platform, we did not filter out the 4053 overlapping repositories from our comparisons between the Penumbra and GitHub.
We again conclude that our Penumbra sample is primarily not captured by Software Heritage; see also Discussion.

We additionally looked for mirrors and forks within the Penumbra, shown in \cref{fig:mirrors}. As when comparing to external datasets, we found repositories that shared a first commit hash, then checked whether the last commit hash diverged. We find 11717 Penumbra repositories share a first commit with at least one other, which constitutes 25.88\% of non-empty Penumbra repositories. These mirrors come from a total of 3348 initial commits. Of these repositories, 6806 share a \textit{last} commit with one or more repositories, suggesting that they have not diverged since creation. Notably, 1287 of the forks and mirrors contain only a single commit. Over a third of the forks and mirrors are on academic hosts (39.46\%, 4623 repositories), which is especially notable because academic hosts constitute only 15\% of our dataset. As a ratio, we find 35.56 mirrors per academic host, 9.98 per non-academic host. This would fit an educational use-case, such as a class assignment where each student clones an initial repository and then works independently.

\section{Discussion}
\label{sec:discussion}

In this paper, we collected independent git hosts to sample what we call the \textit{Penumbra} of the open source ecosystems: public hosts outside of the large, popular, centralized platforms like GitHub.
Our objective was to compare a sample of the Penumbra to GitHub to evaluate the representativeness of GitHub as a data source and identify the potential impact of a platform on the work it hosts.
In doing so, we found that projects outside of centralized platforms were more academic, longer maintained, and more collaborative than those on GitHub.
These conclusions were obtained by looking at domains of email addresses of user accounts in the repositories, as well as measuring temporal and structural patterns of collaborations therein.

Importantly, projects in the Penumbra also appear to be more heterogeneous in important ways.
Namely, we find more skewed distributions of files per repository and average number of editors per file, as well as more bursty patterns of editing.
These bursty patterns are characterized by a skewed distribution of interevent time; meaning, projects in the Penumbra are more likely to feature long periods without edits before periods of rapid editing.
Altogether, our results could suggest that the popularity and very public nature of GitHub might contribute to a large amount of low-involvement contributors (or so-called ``drive-by'' contributions).

Our current sample of the Penumbra is extensive, but our methodology for identifying hosts presents shortcomings. 
Most notably, of the approximately 60,000 GitLab CE, Gitea, and Gogs instances identified by Shodan, only 13.4\% provided public access to one or more repositories. 
We can say little about the hosts that provide no public access, and therefore constitute the dark shadow of software development.
Further, Shodan may not capture all activity on a given server: it identifies hosts by their responses to a request for the front page, and is not a complete web crawler (\cref{subsec:datasets}).
While this was sufficient in identifying 60,000 hosts, it is an under-estimate of the true number of Penumbra hosts, meaning that our dataset remains a sample of the full Penumbra and there exists room for improvement.

We determined from commit hashes that our sample of the Penumbra is mostly disjoint from GitHub and from the Software Heritage archive.
This shows that our strategy of seeking public hosts using Shodan is a viable way to uncover novel sources of code.
Archival efforts such as Software Heritage and World of Code~\cite{world_of_code} can benefit from this work as they can easily integrate our sampling method into their archiving process.
Doing so can help them further achieve their goals of capturing as much open source software as possible.

There remain several open questions about our sample of the Penumbra worth further pursuit. 
For instance, the observed shift in languages used on Penumbra repositories implies that they tend to have more focus on academic and/or scientific projects than GitHub.
However, programming language alone is a coarse signal of the intent and context of a given project. 
Future work should attempt a natural language analysis of repository contents to better identify the type of problems tackled in different regions of the open-source ecosystem. Furthermore, this would allow researchers to match Penumbra and GitHub repositories by the problem-spaces they address, indicating whether developers off of GitHub solve similar problems in different ways.

There are also several important demographic questions regarding, among others, the age, gender, and nationality of users in the Penumbra. 
GitHub is overwhelmingly popular in North America \cite{2020octoverse} and therefore does not provide uniform data on members of the open-source community. 
Critical new efforts could attempt to assess the WEIRDness --- i.e., the focus on Western, educated, industrialized, rich and democratic populations \cite{henrich2010beyond, henrich2010most} --- of GitHub as a convenience sample.

Digging further into the code or user demographics of the Penumbra would allow us to answer new questions about the interplay of code development with the platform that supports it. 
How does the distribution of developer experience levels affect projects, teams and communities? What are the key differences in intent, practices and products based on how open and public the platform is? 
Who contributes to the work and does it differ depending on the platform \cite{casari2021open}? %

We are only beginning to explore the space of open source beyond GitHub and other major central platforms. 
The Penumbra hosts explored here are fundamentally harder to sample and analyze. 
The hosts themselves have to be found and not all hosts provide public access. 
Unlike GitHub, we do not have a convenient API for sampling the digital traces of collaborations, so the underlying git repositories must be analyzed directly. 
There is therefore much of the open source ecosystem left to explore. 
Yet only by exploring new regions, as we did here, can we fully understand how online collaborative work is affected by the platforms and technologies that support it.

\begin{backmatter}

\section*{Availability of data and materials}
The datasets generated during and/or analysed during the current study are available from the corresponding author on reasonable request.

\section*{Competing interests}  
The authors declare that they have no competing interests.
  
\section*{Funding}
All authors were supported by Google Open Source under the Open-Source Complex Ecosystems And Networks (OCEAN) project. 
Any opinions, findings, and conclusions or recommendations expressed in this material are those of the authors and do not necessarily reflect the views of Google Open Source.

\section*{Author's contributions}
M.Z.T.\ conceived of the presented idea, implemented and conducted data collection, and performed data analysis. 
J.B.\ implemented and applied statistical models. 
L.H.D.\ and J.B.\ supervised the project and planned out data collection and analysis.
All authors wrote the final manuscript.

\section*{Acknowledgements}
Computations were performed in part on the Vermont Advanced Computing Core.
\bibliographystyle{spbasic} %
\bibliography{references}      %

\end{backmatter}
\end{document}